\documentstyle[preprint,aps]{revtex}

\draft

\begin{document}


\title{Experimental Studies of Low-field Landau Quantization in Two-dimensional Electron Systems in GaAs/AlGaAs Heterostructures}

\author{Jing-Han Chen$^{1}$, D. R. Hang$^{2,3}$, C. F. Huang$^{4}$, Tsai-Yu Huang$^{1}$, Jyun-Ying Lin$^{1}$, S. H. Lo$^{5}$, J. C. Hsiao$^{4}$, Ming-Gu Lin$^{1}$, M.Y. Simmons$^{6,7}$, D.A. Ritchie$^{6}$, and C.-T. Liang$^{1}$}

\address{$^{1}$Department of Physics, National Taiwan University, Taipei, Taiwan, R.O.C.}

\address{$^{2}$Department of Materials Science and Optoelectronic Engineering, National Sun Yat-sen University, Kaohsiung 804, Taiwan, R.O.C.}

\address{$^{3}$Center for Nanoscience and Nanotechnology,
National Sun Yat-sen University, Kaohsiung 804, Taiwan, R.O.C.
}

\address{$^{4}$National Measurement Laboratory, Center for Measurement Standards, Industrial Technology Research Institute, Hsinchu, Taiwan, R.O.C.}

\address{$^{5}$Institute of Materials Science and Engineering, National Sun Yat-sen University, Kaohsiung 804, Taiwan, R.O.C.}

\address{$^{6}$Cavendish Laboratory, Madingley Road, Cambridge CB3 0HE, United Kingdom}

\address{$^{7}$School of Physics, University of New South Wales, Sydney 2052, Australia}


\maketitle

\begin{abstract}%
By applying a magnetic field perpendicular to GaAs/AlGaAs two-dimensional electron systems, we study the low-field Landau quantization when the thermal damping is reduced with decreasing the temperature. Magneto-oscillations following Shubnikov-de Haas (SdH) formula are observed even when their amplitudes are so large that the deviation to such a formula is expected. Our experimental results show the importance of the positive magneto-resistance to the extension of SdH formula under the damping induced by the disorder.    
\end{abstract}



By applying a magnetic field $B$ perpendicular to the two-dimensional electron systems (2DESs) in semiconductor heterostructures, we can observe the magneto-oscillations in the longitudinal resistivity $\rho _{xx}$ with decreasing the temperature $T$ because of Landau quantization. Such oscillations are expected to follow Shubnikov-de Haas (SdH) formula \cite{Coleridge,SdH}
\begin{eqnarray}
\rho _{xx} \sim \rho _{xx,B=0} + \Delta \rho _{xx} \mathrm{cos} [ \pi ( \nu - 1 ) ]
\end{eqnarray} 
with the oscillating amplitude \cite{GaN_I-QH}
\begin{eqnarray}
\Delta \rho _{xx} = 4 \rho _{0} \frac{2 \pi ^{2} k _{B} T m ^{*} / \hbar e B }{\mathrm{sinh}(2 \pi ^{2} k _{B} T m ^{*} / \hbar e B)} e ^{ - \pi m ^{*} / e \tau _{q} B} 
\end{eqnarray}
at low $B$ when the spin-splitting is unresolved. Here $\nu$ is the filling factor, $\rho _{xx, B=0}$ is the value of $\rho _{xx}$ at $B$=0, $\tau _{q}$ is the quantum lifetime, $k _{B}$ and $m^{*}$ represent the Boltzmann constant and effective mass, $\hbar$ equals Plank constant $h$ divided by $2 \pi$, and $\rho _{0}$ is a constant. It is expected that $\rho _{xx, B=0}$ is independent of the temperature, and the ratio $\rho _{0} / \rho _{xx, B=0} \sim 1$ \cite{Coleridge,GaN_I-QH}. It is well-established how to obtain the carrier concentration $n$, quantum lifetime, and effective mass from the last (oscillating) term of Eq. (1). On the other hand, Landau quantization results in the integer quantum Hall effect (IQHE) \cite{Klitzing} with increasing $B$ when the high-field localization  \cite{high-field-localization} becomes important. Such an effect is characterized by a series of plateaus in the Hall resistivity $\rho _{xy}$ when $\rho _{xx}$ approaches zero. The Hall plateaus are so accurate that the IQHE has been used to maintain the resistance standard. \cite{Delahaye} By studying the magnetic-field-induced transitions in the IQHE, we can investigate the renormalization-group theory \cite{Pruisken} and modular symmetry \cite{Shahar,Dolan}.      
 
The localization arising from the quantum interference is very important to standard theories for the IQHE. \cite{Jiang,Kivelson,Murzin} While these theories are successful at high $B$, they are inappropriate for the low-field Landau quantization in most 2DESs because the localization length may become much longer than the effective size with decreasing $B$. \cite{Jiang} Actually the quantum interference is ignored in SdH formula, which can be obtained from a semiclassical approach. \cite{Coleridge,Coleridge2} Therefore, we shall investigate the low-field Landau quantization to understand the crossover from the semiclassical regime to the IQHE with increasing $B$. To explain the appearance of Hall plateaus when Eq. (1) is still applicable for $\rho _{xx}$, it has been shown that the quantum interference is more robust in $\rho _{xy}$ than that in $\rho _{xx}$ in such a crossover. \cite{Coleridge,Hang} Actually, the IQHE can also be explained by fixing the chemical potential without considering the high-field localization. \cite{Tobias*other*Mahan,Tobias2}. 

According to Eq. (2), the oscillating amplitude $\Delta \rho _{xx}$ increases with increasing the magnetic field $B$. Deviations to SdH formula are expected when the amplitude becomes so large that $\Delta \rho _{xx} \sim \rho _{xx, B=0}$ and thus the minimum of $\rho _{xx}$ approaches 0, which is an important feature of the IQHE. \cite{Coleridge2,Hang} However, our group showed in Ref. \cite{Hang} that Eq. (2) can be applicable even when $\Delta \rho _{xx} > \rho _{xx, B=0}$. The extension of Eq. (2) can be due to the thermal damping \cite{Martin}, and it is important to incorporate the positive magneto-resistance into Eq. (1) for the positivity of $\rho _{xx}$. \cite{Hang} In fact, it has been predicted \cite{Martin} that the extension of SdH formula not only under the thermal damping, but also under the disorder effects determining the Dingle factor $\mathrm{exp} (- \pi m^{*} / e \tau _{q} B)$. To further investigate disorder effects in the low-field Landau quantization, in this paper we probe SdH formula when the thermal damping is reduced by decreasing the temperature $T$. Two GaAs/AlGaAs samples are used for this study. For convenience, we denote them as samples A and B, respectively. The samples are made into Hall patterns by the standard lithography. We study the magneto-transport properties of these two samples by the superconducting magnet and top-loading He$^{3}$ system. Magneto-oscillations following Eq. (2) are observed in these two samples, and the typical IQHE appears with increasing $B$. From SdH oscillations, the carrier concentration $n= 2.8 \times 10 ^{15} \text{ } m^{-2}$ and $ 3.5 \times 10 ^{15} \text{ } m ^{-2}$ for samples A and B, respectively. The scattering mobility $\mu _{c}$ obtained from $\rho _{xx, B=0} = 1/ n e \mu _{c} $ is $5.7 \times 10 ^{2} \text{ } m^{2}/V$-$s$ for sample A and is $44~m ^{2}/V$-$s$ for sample B.   

For convenience, in the following we focus on sample A first. Figure 1 shows the low-field curves of  $\rho _{xx}$ observed in such a sample. A series of oscillations appear in Fig. 1 with increasing the perpendicular magnetic field $B$. The oscillating amplitude should follow
\begin{eqnarray}
\mathrm{ln} \frac{ \Delta \rho _{xx} }{ X / sinh X } = ln  ( 4 \rho _{0}) + \pi m ^{*} / e \tau _{q} B
\end{eqnarray}
with $X=2 \pi ^{2} k _{B} T m ^{*} / \hbar e B $ if Eq. (2) holds true. It is known from the above equation that the curves of $ln \frac{\Delta \rho _{xx}}{X/sinhX}-1/B$ at different temperatures collapse into a straight line in the Dingle diagram if SdH formula is applicable. To avoid effects due to the exchange enhanced spin gaps \cite{Leadley} and asymmetric spin-resolved oscillations \cite{Ando}, we construct such a diagram by considering the spin-degenerate oscillations, which can survive at higher $B$ with increasing $T$. By taking $m^{*}=0.067 m_{0}$ to calculate $X/sinhX$, as shown in Fig. 2, the logarithmic values of $\Delta \rho _{xx} / (X / sinh X) $ for the spin-degenerate oscillations collapse well into a straight line with respect to $1/B$. The slope yields $\mu _{q} = 3.6 \text{ } m ^{2}/ V$-$s$. The good collapse indicates the validity of Eq. (2) and SdH theory. 

The oscillating amplitude $\Delta \rho _{xx}$ at $B>0.35$ T, in fact, is larger than the zero-field longitudinal resistivity $\rho _{xx, B=0}$ with increasing $T$. When $\Delta \rho _{xx} > \rho _{xx, B=0}$, the minimum of $\rho _{xx}$ become negative according to Eq. (1) and thus the deviations to SdH formula is expected. However, Eq. (2) still holds when $B>0.35$ T until the spin-splitting becomes resolved. It has been shown in Ref. \cite{Hang} that the positive magneto-resistance is important to the validity of Eq. (2) when $\Delta \rho _{xx} > \rho _{xx, B=0}$. As shown in the inset of Fig. 2, the positive magnetoresistance becomes apparent after taking the average with respect to the magneto-oscillations to obtain the non-oscillatory background. Hence the experimental results support the importance of the positive magneto-resistance to refine Eq. (1). The extension of Eq. (2) under $\Delta \rho _{xx} > \rho _{xx, B=0}$ in Ref. \cite{Hang} can be due to the thermal-damping factor $X/sinhX$. At $T=0.27 $ K, however, the damping term $X / sinhX > 0.9$ and is close to the zero-temperature value 1 when $B > 0.35$ T. Thus the extension of SdH formula cannot be fully due to the thermal damping in our study. We note that such a formula can also survive when the Dingle factor $\mathrm{exp} (- \pi m^{*} /e \tau _{q} B)$, which represents the disorder, induces strong damping. \cite{Martin} Such a factor is smaller than 0.27 for the spin-degenerate oscillations at $T=0.27$ K, so it can result in the remarkable damping effect. For sample A, the Dingle factor is significant in comparison with the thermal damping when $T<1$ K. Therefore, our observations support importance of the disorder effects to extension of SdH formula when there exists the positive magneto-resistance. 

As mentioned above, we construct the Dingle diagram by only considering the spin-degenerate oscillations to avoid exchange and asymmetric effects. Deviations to SdH theory are expected at the onset of spin splitting although the SdH theory is applicable when such a splitting is fully resolved. \cite{spin-resolved*beating}. At $T=0.27$ K, the effects of the spin-splitting in sample A appear when $B>0.5$ T, where Eq. (2) becomes invalid. With increasing the temperature $T$, the spin-degenerate oscillations can survive at larger $B$ and still follow Eq. (2) when $B=0.5 \sim 1$~T. While the thermal effects in Ref. \cite{Hang} result in the extension of SdH formula by the damping factor $X/sinhX$, our study show that the thermal effect can also suppress the spin-splitting to extend such a formula.            

Refinements to SdH theory are discussed in the literature. The value of $\rho _{0}$, which can be obtained from the the intercept of $\mathrm{ln}\frac{\Delta \rho _{xx}}{X/sinhX}-1/B$ at $1/B \rightarrow 0$ according to Eq. (3), can deviate from $\rho _{xx, B=0}$ although $\rho _{0}/ \rho _{xx, B=0} \sim 1$ is expected in the conventional SdH theory. \cite{GaN_I-QH} In our study, the ratio $ \rho _{0}/ \rho _{xx,B=0} = 3.6$ for sample A. From the reports on high-disorder 2DESs, in fact, it is not always appropriate to relate the constant $\rho _{0}$ to $\rho _{xx, B=0}$ because the zero-field value of $\rho _{xx}$ can depend on the temperature rather than being a constant. \cite{GaN_I-QH}   

A quantum Hall state is characterized by $\rho _{xx} =0$ in addition to the quantized Hall plateau. In sample A, the spin-splitting is resolved before the appearance of the zero longitudinal resistivity as the field $B$ increases. Thus we cannot probe Eq. (2) with well-developed quantum Hall states in such a sample. In sample B, as shown in the inset of Fig. 3, the minimum of $\rho _{xx}$ approach 0 at low temperatures as $B > B _{1}$ when the spin-splitting is still unresolved. Therefore, we can probe SdH formula under the appearance of zero longitudinal resistivity by investigating such a sample. The Dingle diagram of sample B is shown in Fig. 3. By taking $m ^{*} =0.064 \text{ } m _{0} $, we can approximate $\mathrm{ln} \frac{ \Delta \rho _{xx} }{ X/sinhX}-1/B$ by a straight dash line even when $B > B_{1}$. The slope of the straight line yields $\mu _{q} = 2.7 m ^{2}/V$-$s$, and the Dingle term provides stronger damping effects than the thermal factor $X/sinh X$ does at low temperatures as $B>B _{1}$.  Our observations indicate the importance of the disorder effects to the coexistence of quantum Hall states and SdH formula.      

Both the quantum and scattering mobilities of sample B are lower than those of sample A, so the disorder strength should be stronger in the former sample. We note that the disorder may destruct the spin gaps, and hence the spin-splitting is resolved at larger magnetic field in sample B than in sample A. When the minimum of $\rho _{xx} \rightarrow 0$, the oscillating amplitude $\Delta \rho _{xx}$ is determined by the peak values of $\rho _{xx}$. In addition, the non-oscillatory background can be approximated by the half of the envelope function for the peaks of oscillations. Hence the validity of Eq. (2) for sample B when $B>B_{1}$ indicates that SdH formula can provide good estimations to both the peak values and non-oscillatory background of $\rho _{xx}$ under suitable conditions.                                         

In conclusion, the low-field Landau quantization is studied by probing the crossover from semiclassical SdH regime to the integer quantum Hall effect. Our experimental results support the extension of SdH formula, to which we shall include the positive magnetoresistance, under the damping due to the disorder effects. In addition, the thermal effects can suppress the spin-splitting for the extension. When the minimum of $\rho _{xx}$ approaches zero, such a formula may provide estimations to both the peak values and the positive-magnetoresistance background.           
                              
This work is supported by NSC, Taiwan. C.T.L. acknowledges financial support from NSC 94-2112-M-002-037. D.R.H. acknowledges support from NSC 94-2112-M-110-009, ACORC and Aim for the Top University Plan, Taiwan. The work undertaken
at Cambridge was funded by the EPSRC, UK. C.T.L. thanks Tina Liang and Valen Liang for their support.


\centerline{Figure Captions}

Fig. 1 The low-field curves of the longitudinal resistivity of sample A at $T=$0.27, 0.37, 0.47, 0.57, 0.67, 0.82, 0.97, 1.1, 1.3, and 1.4 K, respectively.

Fig. 2 The Dingle diagram for sample A. The straight line is the best fit to $\mathrm{ln} \Delta \rho
 _{xx} - 1/B$. The dash line in the inset shows the non-oscillatory background of the curve of the longitudinal resistivity (the solid line) at $T=0.27$ K.

Fig. 3 The Dingle diagram for sample B. The straight line is the best fitting to $\mathrm{ln} \Delta \rho
 _{xx} - 1/B$. The inset shows the low-field curves of the longitudinal resistivity at different temperatures for sample B.  

\end{document}